\documentclass[10pt,a4paper]{article}

\usepackage[english]{babel}

\usepackage{collref}

\usepackage{amsmath}
\usepackage{amssymb}
\usepackage{graphicx}

\usepackage[colorlinks=true,pdfstartview=FitH]{hyperref} 

\newcommand{\R}{\mathbb{R}} 
\newcommand{\Z}{\mathbb{Z}} 
\DeclareMathOperator{\adj}{adj}
\newcommand{\T}[1]{#1^T} 
\newcommand{\hc}[2][]{#2^{\dagger #1}} 
\newcommand{\abs}[1]{|#1|} 

\newcommand{\figref}[1]{Figure~\ref{#1}}
\newcommand{\secref}[1]{Section~\ref{#1}}

\title{Vacuum Stability Conditions \\ From Copositivity Criteria}
\date{}
\author{Kristjan Kannike}

\begin{document}

\maketitle

\begin{center}
\textit{\footnotesize Scuola Normale Superiore and INFN, Piazza dei Cavalieri 7, 56126 Pisa, Italy\\
National Institute of Chemical Physics and Biophysics, R\"{a}vala 10, Tallinn, Estonia }
\end{center}

\begin{abstract}
A scalar potential of the form $\lambda_{ab} \varphi_{a}^{2} \varphi_{b}^{2}$ is bounded from below if its matrix of quartic couplings $\lambda_{ab}$ is \textit{copositive} -- positive on non-negative vectors. Scalar potentials of this form occur naturally for scalar dark matter stabilised by a $\Z_{2}$ symmetry. Copositivity criteria allow to derive analytic necessary and sufficient vacuum stability conditions for the matrix $\lambda_{ab}$. We review the basic properties of copositive matrices and analytic criteria for copositivity. To illustrate these, we re-derive the vacuum stability conditions for the inert doublet model in a simple way, and derive the vacuum stability conditions for the $\Z_{2}$ complex singlet dark matter, and for the model with both a complex singlet and an inert doublet invariant under a global $U(1)$ symmetry. 
\end{abstract}

\section{Introduction}
\label{sec:introduction}

The Standard Model Higgs potential is not stable up to the scale of Grand Unified Theories (GUT) if the Higgs boson mass is below $128$~GeV, because the Higgs boson quartic coupling $\lambda$ runs to negative values below the GUT scale \cite{Casas:1994qy,Casas:1996aq,Isidori:2001bm,Isidori:2007vm,EliasMiro:2011aa,Buttazzo:2013uya}. Extending the model by scalars can improve vacuum stability \cite{Kadastik:2011aa,Gonderinger:2012rd,EliasMiro:2012ay,Lebedev:2012zw,Chen:2012fa,Cheung:2012nb}. 

To ensure a stable vacuum, the scalar potential has to be bounded from below. In the Standard Model it is enough to have a positive Higgs boson quartic coupling $\lambda > 0$. In theories with more scalar degrees of freedom, such as the two-Higgs-doublet model, the potential should be bounded from below in all directions in the field space as the field strength approaches infinity. 

Finding vacuum stability conditions is a difficult problem in general. However, it is possible that the quartic part of the scalar potential is a quadratic form $\lambda_{ab} \varphi_{a}^{2} \varphi_{b}^{2}$ of squares of real fields. A $\Z_{2}$ symmetry that stabilises scalar dark matter imposes this condition in a natural way. In that case, definite analytic conditions for vacuum stability can be given.

In the calculations of vacuum stability we can ignore any terms with dimensionful couplings---mass terms or soft terms---, since in the limit of large field values, $d < 4$ terms are negligible in comparison with the quartic part of the scalar potential $V_{4}$. Demanding that $V_{4} > 0$ as the fields $\varphi_{i} \to \infty$ is the requirement of strong stability.  $V_{4} \geqslant 0$ means stability in the marginal sense (then one must have $V_{2} \geqslant 0$). Below we give all conditions in the marginal sense; they can be made strong by making the inequalities strict.

If the quartic part of the scalar potential $V_{4}$ is a biquadratic form $\lambda_{ab} \varphi_{a}^{2} \varphi_{b}^{2}$ of real fields or gauge orbit variables, then its domain is not the whole $\R^{n}$, but the non-negative orthant $\R_+^n$. In that case, the potential is positive if the matrix $\lambda_{ab}$ is \emph{copositive}---positive on non-negative vectors (`copositive' is short for `conditionally positive'). 

The notion of copositivity was first introduced by Motz\-kin \cite{Motzkin:1952aa} in 1952. Hitherto it has been used in the field of quadratic optimisation (see e.g. \cite{Quist98copositiverelaxation,springerlink:10.1007/978-3-642-12598-0_1}).
 The set of copositive matrices is larger than the set of familiar positive (semi)definite matrices and includes it. While is easy to find the positive definite part of the parameter space using Sylvester's criterion on $\R^{n}$ (see \secref{sec:copositivity}), the criteria for copositivity are more complex. In general, testing for copositivity is NP-hard \cite{springerlink:10.1007/BF02592948}. Imposing symmetries on the potential can considerably simplify the criteria.

While the notion of copositivity is well known in the field of optimisation and nonlinear programming, to our knowledge we are the first to use these criteria to determine constraints on quartic couplings from vacuum stability. Recent reviews on copositive matrices are \cite{springerlink:10.1007/BF02355379,Hiriart-Urruty:2010fk} and in the Ph.D. thesis of Bundfuss \cite{Bundfuss:2009fk}. A short review of many specific algorithms to detect copositivity is \cite{DBLP:journals/eor/Bomze12}. 

The purpose of this paper is to introduce basic properties of copositive matrices and criteria for copositivity. We compare copositivity with the familiar positive definiteness. We give explicit criteria for copositivity of $2 \times 2$ and $3 \times 3$ matrices, and some general criteria in terms of the cofactors of the matrix or in terms of its eigenvalues and eigenvectors. We have left out lengthy proofs of the theorems we give and refer the interested reader to the cited papers.

After a review we put the criteria to practice. We re-derive the vacuum stability conditions for the inert doublet model \cite{Deshpande:1977rw,Barbieri:2006dq,Ma:2006km,LopezHonorez:2006gr} in a parameterisation where they are self-evident. Then we derive the necessary and sufficient vacuum stability conditions \eqref{eq:SM:singlet:copos:crit} for the full scalar potential of the Higgs boson and  complex singlet dark matter stabilised by a $\Z_{2}$ symmetry. Previously, only a simpler potential with a global $U(1)$ has been considered \cite{McDonald:1993ex,Barger:2008jx}. We see that in case of singlets, in general their real components, not the polar form, have to be used to capture the whole parameter space. Last, we derive the vacuum stability conditions for a model where the scalar dark sector comprises both the inert doublet and a complex singlet \cite{Kadastik:2009dj,Kadastik:2009cu}.

In \secref{sec:copositivity}, we introduce the notion of copositivity, give some criteria for matrices to be copositive, and compare them to the usual criteria of positive definiteness. In \secref{sec:SM:ID}, we re-derive the vacuum stability conditions for the inert doublet model. In \secref{sec:SM:complex:singlet}, we derive the vacuum stability conditions for the model of $\Z_{2}$ complex singlet dark matter. To give another example, we find vacuum stability conditions for the singlet and doublet model with a global $U(1)$ symmetry in \secref{sec:U:1:SIID}. We conclude in \secref{sec:conclusions}.

\section{Copositive Matrices}
\label{sec:copositivity}

We first briefly recall properties of the usual positive matrices. \emph{A symmetric matrix $A$ is said to be \emph{positive semidefinite} if the quadratic form $\T{x} A x \geqslant 0$ for all vectors $x$ in $\R^n$. -- A positive matrix $A$ is \emph{positive definite} if the quadratic form $\T{x} A x > 0$ for all vectors $x$.}

There are several equivalent criteria for a matrix $A$ to be positive. For example, all its eigenvalues have to be non-negative, as we can go to a basis where the matrix $A$ is diagonal. If all $x_i$ save one are zero, then in that basis $a_{ii} \geqslant 0$ for the matrix to be positive.  For practical purposes, however, this is not a good criterion: already the eigenvalues of a $3 \times 3$ matrix are rather complicated analytical expressions.

A more convenient one is Sylvester's criterion (see e.g. \cite{Gilber:1991:PDM:115430.115439}). It states that for a symmetric matrix $A$ to be positive semidefinite, the principal minors of $A$ have to be non-negative. (The principal minors are determinants of the principal submatrices. The principal submatrices of $A$ are obtained by deleting rows and columns of $A$ in a symmetric way, i.e. if the $i_{1}, \ldots, i_{k}$ rows are deleted, then the $i_{1}, \ldots, i_{k}$ columns are deleted as well. The largest principal submatrix of $A$ is $A$ itself.) Thus if the matrix $A$ is positive, all of its submatrices, in particular the diagonal elements $a_{ii}$ have to be non-negative.

The definition of a copositive matrix is similar to the definition of a positive definite matrix.
\emph{A symmetric matrix $A$ is \emph{copositive} if the quadratic form $\T{x} A x \geqslant 0$ for all vectors $x \geqslant 0$ in the non-negative orthant $\R_+^n$.} (Notation $x \geqslant 0$ means that $x_i \geqslant 0$ for each $i = 0, \ldots, n$.) -- \emph{A copositive matrix $A$ is \emph{strictly copositive} if the quadratic form $\T{x} A x > 0$ for all vectors $x > 0$.}

Positive matrices are a subset of copositive matrices. It is also obvious that a non-negative matrix $A$ with $a_{ij} \geqslant 0$ of any order is copositive. It is easy to show that for $n = 2$ copositive matrix is either positive-semidefinite or non-negative. In fact every copositive matrix of order $n = 3$ or $n = 4$ can be expressed as the sum  $S+N$ of a positive semidefinite matrix $S$ and a nonnegative matrix $N$ \cite{CambridgeJournals:2051276}. For $n > 4$, matrices of the form $S+N$ are a strict subset of the copositive matrices.%
\footnote{Thus random generation of copositive matrices is rather easy for $n \leq 4$, since by Cholesky decomposition every positive semidefinite matrix can be factorised as $S = L \hc{L}$, where $L$ is a lower triangular matrix.} 
However, because for large matrices the general tests are slow, it can still be useful to test if a given matrix has the form $S+N$ \cite{Kaplan2001237}.
 
If one component $x_{i}$ is set to zero, then the quadratic form $\T{x} A x$ becomes a quadratic form of the remaining variables. Therefore if a matrix $A$ of order $n$ is copositive, each principal submatrix of $A$ of order $n-1$ is also copositive and so forth. In particular $a_{ii} \geqslant 0$ for all $i$. If $A$ is strictly copositive then $a_{ii} > 0$.

A symmetric matrix $A$ of order 2 is copositive if and only if \cite{KP198379}
\begin{align}
  a_{11} \geqslant 0, a_{22} &\geqslant 0, \label{eq:A:2:copos:diag:1:2}
  \\
  a_{12} + \sqrt{ a_{11} a_{22} } &\geqslant 0. \label{eq:A:2:copos:nondiag:12}
\end{align}
This can be easily seen: if $a_{12} < 0$, then $a_{11} a_{22} \geqslant a_{12}^{2}$ is required, but if $a_{12} \geqslant 0$, there are no constraints on $a_{11}$ and $a_{22}$ besides being positive: $A$ is indeed either positive semidefinite or non-negative. Combining these statements with $a_{11} \geqslant 0, a_{22} \geqslant 0$ gives \eqref{eq:A:2:copos:nondiag:12}.

Because \eqref{eq:A:2:copos:nondiag:12} has to hold for all $2 \times 2$ principal submatrices of $A$, we have $a_{ij} + \sqrt{a_{ii} a_{jj}} \geqslant 0$ for all $i,j$. If $a_{ii} = 0$, then $a_{ij} \geqslant 0$ for all $j$.

Note that unlike positive definiteness, copositivity is not invariant under basis transformations. Consider \cite{Bundfuss:2009fk}
\begin{equation}
  A =
  \begin{pmatrix}
    1 & 2 \\
    2 & 1
  \end{pmatrix}
  \text{ and }
  U =
  \begin{pmatrix}
    0 & -1 \\
    1 & 0
  \end{pmatrix}.
\end{equation}
The matrix
\begin{equation}
  \T{U} A U = 
  \begin{pmatrix}
  1 & -2 \\
  -2 & 1
  \end{pmatrix}
\end{equation}
is not copositive as it does not satisfy \eqref{eq:A:2:copos:nondiag:12}. However, copositivity is invariant under permutations and scaling with positive numbers.

A symmetric matrix $A$ of order 3 is copositive if and only if \cite{Hadeler198379,Chang1994113}
\begin{equation}
\begin{split}
  a_{11} &\geqslant 0, a_{22} \geqslant 0, a_{33} \geqslant 0,
  \\
  \bar{a}_{12} &= a_{12} + \sqrt{ a_{11} a_{22} } \geqslant 0,
  \\
  \bar{a}_{13} &= a_{13} + \sqrt{ a_{11} a_{33} } \geqslant 0,
  \\
  \bar{a}_{23} &= a_{23} + \sqrt{ a_{22} a_{33} } \geqslant 0,
\end{split}
\label{eq:A:3:copos:1:2:crit}
\end{equation}
and
\begin{equation}
  \sqrt{a_{11} a_{22} a_{33}} + a_{12} \sqrt{a_{33}} + a_{13} \sqrt{a_{22}} + a_{23} \sqrt{a_{11}} 
  + \sqrt{2 \bar{a}_{12} \bar{a}_{13} \bar{a}_{23}} \geqslant 0.
\label{eq:A:3:copos:3:crit}
\end{equation}
The last criterion in \eqref{eq:A:3:copos:3:crit} is a simplification of
\begin{align}
  \sqrt{a_{11} a_{22} a_{33}} 
  + a_{12} \sqrt{a_{33}} + a_{13} \sqrt{a_{22}} + a_{23} \sqrt{a_{11}} &\geqslant 0,
  \label{eq:A:3:copos:3:crit:new}
  \\
  \det A = a_{11} a_{22} a_{33} 
  - (a_{12}^2 a_{33} + a_{13}^2 a_{22} + a_{11} a_{23}^2) 
  + 2 a_{12} a_{13} a_{23} &\geqslant 0,
  \label{eq:A:3:copos:3:crit:det}
\end{align}
where one \emph{or} the other inequality has to hold \cite{Hadeler198379}. (The condition $\det A \geqslant 0$ is also part of Sylvester's criterion for positive semidefiniteness.)
The conditions \eqref{eq:A:3:copos:1:2:crit} simply state that the three $2 \times 2$ principal submatrices of $A$ are copositive. Only \eqref{eq:A:3:copos:3:crit} looks more peculiar.


In the literature there are also explicit criteria for matrices of order 4 and 5 \cite{Ping1993109} and \cite{Andersson19959} (also gives a recursive algorithm for matrices of order $n$), but these are too lengthy to reproduce here. 

For larger matrices, it is easier to use more general criteria such as the Cottle-Habetler-Lemke theorem \cite{Cottle1970295}: Suppose that the order $n-1$ principal submatrices of a real symmetric matrix $A$ of order $n$ are copositive. In that case A is \emph{non}-copositive if and only if 
\begin{equation}
  \det A < 0 \quad \text{and} \quad \adj A \geqslant 0.
\end{equation}
The adjugate of $A$ is the transpose of the cofactor matrix of $A$:
\begin{equation}
(\adj A)_{ij} = (-1)^{i+j} M_{ji},
\end{equation}
where $M_{ij}$ is the $(i,j)$ minor of $A$, the determinant of the submatrix that results from deleting row $i$ and column $j$ of $A$. 

To illustrate the theorem, consider the $2 \times 2$ symmetric matrix
\begin{equation}
  A_{2} = 
  \begin{pmatrix}
    a_{11} & a_{12} \\
   a_{12} & a_{22} \\
  \end{pmatrix}.
\end{equation}
Its adjugate is
\begin{equation}
  \adj A_{2} = 
  \begin{pmatrix}
    a_{22} & -a_{12} \\
   -a_{12} & a_{11} \\
  \end{pmatrix}.
\end{equation}
By assumption, $a_{11} \geqslant 0, a_{22} \geqslant 0$. The first condition $\det A_{2} = a_{11} a_{22} - a_{12}^2 < 0$ gives by 
\begin{equation}
  a - b^2 \geqslant 0 \Rightarrow \sqrt{a} + b \geqslant 0
  \label{eq:sq:to:sqrt:lemma}
\end{equation}
that if $a_{12} + \sqrt{a_{11} a_{12}} < 0$, then $A$ is not copositive. Indeed, this violates \eqref{eq:A:2:copos:nondiag:12}. For the $2 \times 2$ matrix the second condition does not give anything new because $a_{12}$ may or may not be negative without  violating copositivity.

The theorem can be rephrased: Suppose that the order $n-1$ principal submatrices of a real symmetric matrix $A$ of order $n$ are copositive. In that case A is copositive if and only if 
\begin{equation}
  \det A \geqslant 0 \quad \text{or\quad some element(s) of} \adj A < 0.
\end{equation}

For $A_{2}$, the condition $\det A_{2} \geqslant 0$ together with $a_{11} \geqslant 0, a_{22} \geqslant 0$ is Sylvester's criterion of positive semidefiniteness for $2 \times 2$ matrices. Alternatively some elements of $\adj A_{2} < 0$. Because $a_{11} \geqslant 0, a_{22} \geqslant 0$, it can be only  $-a_{12} < 0$. The condition $a_{11} a_{22} - a_{12}^2 \geqslant 0 \vee a_{12} > 0$ is exactly equivalent to \eqref{eq:A:2:copos:nondiag:12}.


There are other algebraic criteria for copositivity. Positive definiteness can be checked via positivity of eigenvalues of the matrix. A similar criterion exists for copositivity, only now all the eigenvalues and eigenvectors of the principal submatrices have to be considered as well.  -- Kaplan's test \cite{Kaplan:2000:TCM} states that a symmetric matrix  $A$ is copositive if and only if every principal submatrix of $A$ has no eigenvector $v > 0$ with associated eigenvalue $\lambda < 0$. (This arises from the positivity of the so-called Pareto eigenvalues of the matrix $A$ \cite{Hiriart-Urruty:2010fk}.) 

There is also a Schur complement-like theorem for copositive matrices \cite{Bomze:1989fk,Bomze1996161} stating that the matrix
\begin{equation}
  A = 
  \begin{pmatrix}
  a & \T{b} \\
  b & C
  \end{pmatrix},
\end{equation}
where $b$ is a vector in $\R^{n-1}$ and $C$ is a symmetric matrix of order $n$, is copositive if and only if
\begin{enumerate}
\item $a \geqslant 0$, $C$ is copositive, and
\item $\T{y} (a C - b \T{b}) y \geqslant 0$ for all $y \in \R^{n-1}$ such that $\T{b} y \leqslant 0$. 
\end{enumerate}
The second condition is the hardest to verify.

Because the number of principal submatrices grows exponentially with matrix size, for large matrices of order about $n > 20$ numerical criteria like simplex algorithms \cite{MR2388635} should be used.

\section{An Example: Vacuum Stability Conditions for the Inert Doublet Model}
\label{sec:SM:ID}

For a warmup, we re-derive the vacuum stability conditions for the inert doublet model \cite{Deshpande:1977rw,Barbieri:2006dq,Ma:2006km,LopezHonorez:2006gr} via copositivity. The inert doublet model is a $\Z_{2}$-symmetric version of the two-Higgs-doublet model (2HDM) \cite{Lee:1973iz,Branco:1980sz} (see \cite{Branco:2011iw} for a comprehensive review) that contains a candidate of scalar dark matter. For completeness we also discuss the general 2HDM, though it is more amenable to other approaches.

The most general 2HDM scalar potential of two electroweak doublets $H_{1}$ and $H_{2}$ is
\begin{equation}
\begin{split}
  V &= \mu_{1}^{2} \abs{H_{1}}^{2} + \mu_{2}^{2} \abs{H_{2}}^{2} 
  + \lambda_{1} \abs{H_{1}}^{4} + \lambda_{2} \abs{H_{2}}^{4} 
  + \lambda_{3} \abs{H_{1}}^{2} \abs{H_{2}}^{2} \\
  &+ \lambda_{4} (\hc{H_{1}} H_{2}) (\hc{H_{2}} H_{1}) 
  + \frac{1}{2} \left[ \lambda_{5} (\hc{H_{1}} H_{2})^{2} + \lambda_{5}^{*} (\hc{H_{2}} H_{1})^{2} \right] \\
  &+ \abs{H_{1}}^{2} (\lambda_{6} \hc{H_{1}} H_{2} + \lambda_{6}^{*} \hc{H_{2}} H_{1})
  + \abs{H_{2}}^{2} (\lambda_{7} \hc{H_{1}} H_{2} + \lambda_{7}^{*} \hc{H_{2}} H_{1})
   \\
  & = \mu_{1}^{2} h_{1}^{2} + \mu_{2}^{2} h_{2}^{2} 
  + \lambda_{1} h_{1}^{4} + \lambda_{2} h_{2}^{4} 
  + \lambda_{3} h_{1}^{2} h_{2}^{2} + \lambda_{4} \rho^{2} h_{1}^{2} h_{2}^{2} \\
  &+ \abs{\lambda_{5}} \cos(2 \phi + \phi_{\lambda_5}) \rho^{2} h_{1}^{2} h_{2}^{2}
  + 2 r_{1}^{3} r_{2} \abs{\lambda_{6}} \cos(\phi + \phi_{\lambda_{6}}) \\
  &+ 2 r_{1} r_{2}^{3} \abs{\lambda_{7}} \cos(\phi + \phi_{\lambda_{7}}),
\end{split}
\label{eq:V:2HDM}
\end{equation}
where we have parameterised the field bilinears as \cite{Ginzburg:2004vp}
\begin{equation}
  \abs{H_{1}}^{2} = h_{1}^{2}, \quad \abs{H_{2}}^{2} = h_{2}^{2}, 
  \quad \hc{H_{1}} H_{2} = h_{1} h_{2} \rho e^{i \phi}.
  \label{eq:V:2HDM:param}
\end{equation}
The parameter $\abs{\rho} \in [0,1]$ as implied by the Cauchy inequality $0 \leqslant \abs{\hc{H_{1}} H_{2}} \leqslant \abs{H_{1}} \abs{H_{2}}$.

In the inert doublet model, the potential is required to be invariant under the $\Z_{2}$ transformations $H_1 \to H_1$, $H_{2} \to -H_{2}$, setting $\lambda_{6} = \lambda_{7} = 0$. In this case the quartic part of the scalar potential is a quadratic form of $h_{1}^{2}$ and $h_{2}^{2}$. To find the vacuum stability conditions, we must minimise the potential with respect to the free parameters $\rho$ and $\phi$. It is obvious that the minimum of the $\lambda_{5}$ term is given by $\cos(2 \phi + \phi_{\lambda_5}) = -1$. For this minimum, the matrix of quartic couplings for the minimal potential in the $(h_{1}^{2}, h_{2}^{2})$ basis is 
\begin{equation}
 \Lambda 
 =
 \begin{pmatrix}
    \lambda_{1} 
    & 
    \frac{1}{2} \left[ \lambda_{3} 
    + \rho^{2} \left( \lambda_{4} - \abs{\lambda_{5}} \right) \right]
    \\ 
    \frac{1}{2} \left[ \lambda_{3} 
    + \rho^{2} \left( \lambda_{4} -\abs{\lambda_{5}} \right) \right] 
    & 
    \lambda_{2}
   \end{pmatrix}.
    \label{eq:lambdas:SM:ID}
\end{equation}
Now if $\lambda_{4} - \abs{\lambda_{5}} \geqslant 0$, the minimum of the potential is obtained by setting $\rho = 0$, but if $\lambda_{4} - \abs{\lambda_{5}} < 0$, then the minimum is given by $\rho = 1$. 
Applying the copositivity criteria \eqref{eq:A:2:copos:diag:1:2}, \eqref{eq:A:2:copos:nondiag:12} to \eqref{eq:lambdas:SM:ID} for these two cases yields
\begin{equation}
  \lambda_{1} \geqslant 0, \quad \lambda_{2} \geqslant 0, 
  \quad \lambda_{3} + 2 \sqrt{\lambda_{1} \lambda_{2}} \geqslant 0,
  \quad \lambda_{3} + \lambda_{4} - \abs{\lambda_{5}} + 2 \sqrt{\lambda_{1} \lambda_{2}} \geqslant 0,
  \label{eq:SM:ID:copos:crit}
\end{equation}
the well known vacuum stability conditions \cite{Deshpande:1977rw,Klimenko:1984qx,Nie:1998yn,Kanemura:1999xf,Ginzburg:2004vp,Maniatis:2006fs,Ivanov:2006yq} for the inert doublet model.%
\footnote{Had we attempted to find the vacuum stability conditions from copositivity via real components of the fields, we would get, besides \eqref{eq:SM:ID:copos:crit}, redundant inequalities like $\lambda_{2} + \abs{\lambda_{2}} \geqslant 0$, due to the $SU(2)$ symmetry. In the basis of the eight real component fields, it would be much more troublesome to show that the conditions \eqref{eq:SM:ID:copos:crit} are not just necessary, but also sufficient.}

Derivation of the necessary and sufficient conditions \eqref{eq:SM:ID:copos:crit} for the special case of the inert doublet model is simpler with copositivity than with the methods used to derive the necessary and sufficient positivity conditions for the most general 2HDM \cite{Maniatis:2006fs,Ivanov:2006yq}. 

To apply copositivity to the latter, one would have to consider a larger monomial basis $\vec{h} \equiv (h_{1}^{2}, h_{1} h_{2}, h_{2}^{2})$ in which the matrix of couplings is
\begin{equation}
\Lambda (c_{i})
 =
 \begin{pmatrix}
    \lambda_{1}
    &
    \abs{\lambda_{6}} \rho \cos(\phi + \phi_{\lambda_{6}})
    & 
    c_{1} \frac{1}{2} \lambda_{345}(\rho,\phi)
    \\ 
    \abs{\lambda_{6}} \rho \cos(\phi + \phi_{\lambda_{6}})
    &
    c_{2} \lambda_{345}(\rho,\phi)
    & 
    \abs{\lambda_{7}} \rho \cos(\phi + \phi_{\lambda_{7}})
    \\
    c_{1} \frac{1}{2} \lambda_{345}(\rho,\phi)
    & 
    \abs{\lambda_{7}} \rho \cos(\phi + \phi_{\lambda_{7}})
    &
    \lambda_{2}
   \end{pmatrix},
    \label{eq:lambdas:SM:2HDM}
\end{equation}
where we have denoted $\lambda_{345}(\rho,\phi) \equiv  \lambda_{3} 
    + \rho^{2} \left[ \lambda_{4} 
    + \abs{\lambda_{5}} \cos(2 \phi + \phi_{\lambda_{5}}) \right] $.
    
Now, however, the elements on the minor diagonal of the matrix \eqref{eq:lambdas:SM:2HDM} are not independent any more \cite{springerlink:10.1007/s10107-003-0387-5}: they depend on two parameters $c_{1}$ and $c_{2}$ with the condition that 
\begin{equation}
  c_{1} + c_{2} = 1.
  \label{eq:c1:c2:1}
\end{equation}
The matrices $\Lambda (c_{i})$ form an affine space.

It is easy to see that if $\Lambda (c_{i})$ is copositive for some values of $c_{i} = c_{i}^{0}$ satisfying \eqref{eq:c1:c2:1} (and the whole ranges of $\rho$ and $\phi$), then the potential \eqref{eq:V:2HDM} is bounded below. (Note that $c_{i}^{0}$ depend on the values of $\lambda_{i}$, $\rho$ and $\phi$.) 

Namely, assuming that $\Lambda (c_{i}^{0})$ is copositive, but the potential is not bounded below for a non-negative vector $\vec{h}_{-}$ is contradictory, because the potential \eqref{eq:V:2HDM} is given by $V(\vec{h}) = \T{\vec{h}} \Lambda (c_{i}) \vec{h}$ for \emph{any} allowed values of $c_{i}$, especially $c_{i}^{0}$. Thus also for $\vec{h}_{-}$, we have $V(\vec{h}_{-}) = \T{\vec{h}_{-}} \Lambda (c_{i}^{0}) \vec{h}_{-} \geqslant 0$ from non-negativity of $\vec{h}_{-}$ and definition of copositivity.

\section{Vacuum Stability Conditions for Complex Singlet Dark Matter}
\label{sec:SM:complex:singlet}

A real \cite{Silveira:1985rk,Burgess:2000yq,Barger:2007im,Gonderinger:2009jp,Chen:2012fa,Cheung:2012nb} or complex \cite{McDonald:1993ex,Barger:2008jx,Barger:2010yn,Gonderinger:2012rd} scalar singlet is perhaps the simplest candidate for dark matter. The scalar sector of the latter model comprises the Standard Model Higgs doublet $H_{1}$ and the complex singlet
\begin{equation}
  S = \frac{S_H + i \, S_A}{\sqrt{2}}.
\end{equation}

The most general scalar potential invariant under the $\Z_{2}$ transformations $H_1 \to H_1$, $S \to -S$ is
\begin{equation}
\begin{split}
    V &= \mu_{1}^{2} \abs{H_{1}}^{2} + \mu_{S}^{2} \abs{S}^{2} 
    + \frac{\mu_{S}^{\prime 2}}{2} \left[ S^{2} + (\hc{S})^{2} \right]
    + \lambda_{1} \abs{H_{1}}^{4}
    \\
    &+ \lambda_{S} \abs{S}^{4} + \frac{ \lambda'_{S} }{2} \left[ S^{4} 
    + (\hc{S})^{4} \right] 
    + \frac{ \lambda''_{S} }{2} \abs{S}^{2} \left[ S^{2} + (\hc{S})^{2} \right] \\ 
    &+ \lambda_{S1} \abs{S}^{2} \abs{H_{1}}^{2}
    + \frac{ \lambda'_{S1} }{2} \abs{H_{1}}^{2} \left[ S^{2} + (\hc{S})^{2} \right].
\end{split}
\label{eq:V:SM:singlet}
\end{equation}

Only the Hermitian square $\abs{H_{1}}^{2}$ of the Higgs field appears since it is the only scalar doublet in the model, and only the squares $S_H^2$ and $S_A^2$ of the components of the singlet appear because of the $\Z_{2}$ symmetry.

The matrix of quartic couplings $\Lambda$ in the $(h_{1}^{2}, S_{H}^{2}, S_{A}^{2})$ basis is given by
\begin{equation}
  4 \, \Lambda 
  =
  \begin{pmatrix}
    4 \lambda_{1} & \lambda_{S1} + \lambda'_{S1} & \lambda_{S1} - \lambda'_{S1} 
    \\
    \lambda_{S1} + \lambda'_{S1} & \lambda_{S} + \lambda'_{S} + \lambda''_{S} 
    & \lambda_{S} - 3 \lambda'_{S}
    \\
    \lambda_{S1} - \lambda'_{S1}  & \lambda_{S} - 3 \lambda'_{S} 
    & \lambda_{S} + \lambda'_{S} - \lambda''_{S}
  \end{pmatrix}.
    \label{eq:lambdas:SM:singlet}
\end{equation}

Applying the criteria for copositivity \eqref{eq:A:3:copos:1:2:crit} and \eqref{eq:A:3:copos:3:crit} to the matrix of quartic couplings \eqref{eq:lambdas:SM:singlet} yields

\begin{equation}
\begin{split}
  &\lambda_{1} \geqslant 0, \quad 
  \lambda_S + \lambda'_S + \lambda''_S \geqslant 0, \quad 
  \lambda_S + \lambda'_S - \lambda''_S \geqslant 0,
  \\
  &\bar{\lambda}_{1H} = \lambda_{S1} + \lambda'_{S1} 
  + 2 \sqrt{ \lambda_{1} (\lambda_{S} + \lambda'_{S} + \lambda''_{S}) } \geqslant 0,
  \\
  &\bar{\lambda}_{1A} = \lambda_{S1} - \lambda'_{S1} 
  + 2 \sqrt{ \lambda_{1} (\lambda_{S} + \lambda'_{S} - \lambda''_{S}) } \geqslant 0,
  \\
  &\bar{\lambda}_{HA} = \lambda_{S} - 3 \lambda'_{S} 
  + \sqrt{ (\lambda_S+\lambda'_S)^2-\lambda_S^{\prime \prime 2} } 
  \geqslant 0,
  \\
  &2 \sqrt{\lambda_1} \left[ \sqrt{ (\lambda_S+\lambda'_S)^2-\lambda_S^{\prime \prime 2}}
  + \lambda_S - 3 \lambda'_S \right] \\
  &+ \lambda _{S1} 
  \left (\sqrt{\lambda _S+\lambda'_S-\lambda''_S}+\sqrt{\lambda _S+\lambda'_S+\lambda''_S} \right) \\
  &+ \lambda'_{S1} \left( \sqrt{\lambda _S+\lambda'_S-\lambda''_S} 
  - \sqrt{\lambda _S+\lambda'_S+\lambda''_S} \right) \\ 
  &+ \sqrt{2} \sqrt{\bar{\lambda}_{1H}  \bar{\lambda}_{1A} \bar{\lambda}_{HA}} 
  \geqslant 0.
\end{split}
\label{eq:SM:singlet:copos:crit}
\end{equation}
We emphasise that these are the necessary and sufficient conditions for vacuum stability.%
\footnote{The previously given conditions in \cite{Kadastik:2009cu} pertaining to $H_1$ and $S$ are necessary conditions for positivity, not copositivity.}

Previously \cite{Barger:2008jx}, the vacuum stability conditions have been derived only in the case of an extra global $U(1)$ symmetry imposed on  the potential \eqref{eq:V:SM:singlet}. Then $\lambda'_S = \lambda''_S = \lambda'_{S1} = 0$, and the conditions \eqref{eq:SM:singlet:copos:crit} reduce to
\begin{equation}
  \lambda_1 \geqslant 0, \quad \lambda_S \geqslant 0, \quad \lambda_{S1} \geqslant -2 \sqrt{\lambda_1 \lambda_S}
  \label{eq:SM:singlet:U1:copos:crit}
\end{equation}
that -- taking into account normalisation conventions -- coincide with the conditions given in \cite{Barger:2008jx}.
In that case the last inequality in \eqref{eq:SM:singlet:copos:crit} becomes redundant, being identically true by virtue of \eqref{eq:SM:singlet:U1:copos:crit}.

Could one derive the conditions \eqref{eq:SM:singlet:copos:crit} by parameterising the singlet field as $S = s e^{i \phi_{S}}$ and considering the copositivity criteria \eqref{eq:A:2:copos:diag:1:2} and \eqref{eq:A:2:copos:nondiag:12} for the $2 \times 2$ matrix
\begin{equation}
  \Lambda = 
  \begin{pmatrix}
  \lambda_{1} 
  & \frac{1}{2} (\lambda_{S1} + \lambda'_{S1} \cos 2 \phi_{S})
  \\
  \frac{1}{2} (\lambda_{S1} + \lambda'_{S1} \cos 2 \phi_{S}) 
  & \lambda_{S} + \lambda'_{S} \cos 4 \phi_{S} + \lambda''_{S} \cos2 \phi_{S}
  \end{pmatrix}
  \label{eq:lambdas:SM:singlet:modulus}
\end{equation}
of couplings?

In the $U(1)$ case we just encountered, the terms dependent on $\phi_{S}$ are forbidden and \eqref{eq:lambdas:SM:singlet:modulus} gives exactly the conditions \eqref{eq:SM:singlet:U1:copos:crit}.

Otherwise, we have to minimise the potential with respect to $\phi_{S}$. If $h_{1} = 0$, we must have
\begin{equation}
  \lambda_S + \lambda'_S \cos 4 \phi_S + \lambda''_S \cos 2 \phi_S \geqslant 0.
\end{equation}
If $\lambda''_S = 0$, then the minimum of $\lambda_S + \lambda'_S \cos 4 \phi_S$ is obviously $\lambda_S - \abs{\lambda'_S}$. 
Else, the extremum condition is 
\begin{equation}
  2 \lambda'_S \sin 4 \phi_S + \lambda''_S \sin 2 \phi_S 
  = (\lambda'_S + 4 \lambda''_S \cos 2 \phi_S) \sin 2 \phi_S = 0,
\end{equation}
giving $\phi_S = \pm n \frac{\pi}{2}$ and 
$\phi_S = \frac{1}{2} \left[ \pm \arccos \left(-\frac{\lambda''_S}{4 \lambda'_S} \right) + 2 n \pi \right]$.
The former solution reproduces 
\begin{equation}
\lambda_S + \lambda'_S \pm \lambda''_S \geqslant 0,  
\label{eq:S:self-couplings:copos:min:always}
\end{equation} 
whereas the latter solution gives
\begin{equation}
  \lambda_{S} - \lambda'_{S} - \frac{\lambda_{S}^{\prime\prime 2}}{8 \lambda'_{S}} \geqslant 0.
  \label{eq:S:self-couplings:copos:min}
\end{equation}
that only has to hold if the argument of the arccosine is within
\begin{equation}
 -1 \leqslant -\frac{\lambda''_S}{4 \lambda'_S} \leqslant 1.
 \label{eq:arccos:cond}
\end{equation}

From Sylvester's criterion for the $23$ submatrix of \eqref{eq:lambdas:SM:singlet} one gets a similar but more restrictive condition for the usual positivity of singlet self-couplings:
\begin{equation}
  8 (\lambda_{S} - \lambda'_{S}) \lambda'_{S} - \lambda_{S}^{\prime\prime 2} \geqslant 0.
  \label{eq:S:self-couplings:pos}
\end{equation}

\begin{figure}[t]
\centering
\includegraphics{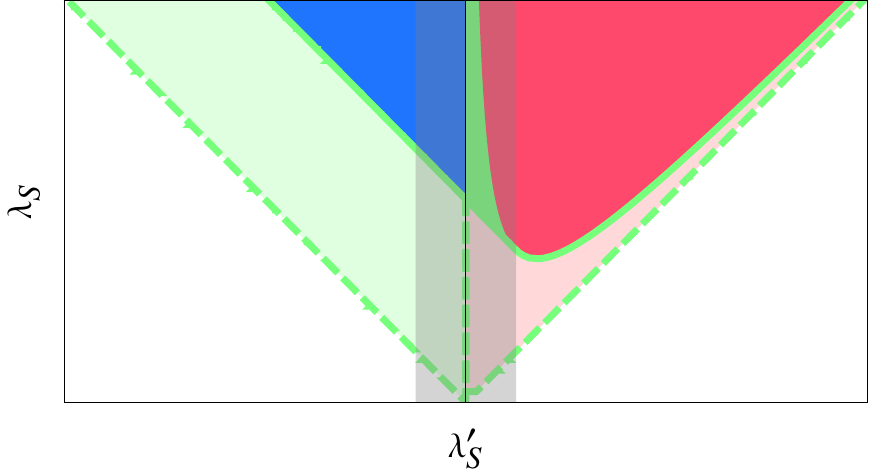}
\caption{Allowed parameter space for the quartic self-couplings of $S$. Area with dashed border: the parameter space allowed for $\lambda''_S = 0$. Only $\lambda'_S \geqslant 0$ is allowed by positivity (light red), while copositivity also includes the area below zero (light green). Area with solid border: the parameter space allowed by positivity (red); the constraint from copositivity (the whole area) for $\lambda''_S = 1/2$. In the transparent grey area \eqref{eq:arccos:cond} and thus \eqref{eq:S:self-couplings:copos:min} do not hold; if we did not take that into account, the green sliver would be erroneously excluded.} 
\label{fig:S:self-couplings}
\end{figure}

The regions are illustrated in \figref{fig:S:self-couplings}. If $\lambda''_{S} = 0$ (area with dashed border), positivity only allows $\lambda_{S} \geqslant \lambda'_{S} \geqslant 0$ (light red), while copositivity allows $\lambda_{S} - \abs{\lambda'_{S}} \geqslant 0$ (light red and light green).

For $\lambda''_{S} = 1/2$, the positivity condition \eqref{eq:S:self-couplings:pos} allows only the red region, the copositivity conditions \eqref{eq:S:self-couplings:copos:min:always} and \eqref{eq:S:self-couplings:copos:min} with minimising the potential allows the area bounded by solid green line, comprising the red, green and blue regions (this is equivalent to $\bar{\lambda}_{HA} \geqslant 0$ \eqref{eq:SM:singlet:copos:crit}). Note that if the condition \eqref{eq:arccos:cond} were not taken into account, the green wedge would be erroneously excluded.

We note that while it \emph{is} possible to use the singlet in polar form, it is tricky already in the simple case of singlet self-couplings. More complications arise when the full set of conditions is considered; in this case, each inequality has to be minimised separately with respect to $\phi_S$, because they have to hold true for its whole range.

\section{Complex Singlet and Inert Doublet with a Global $U(1)$ Symmetry}
\label{sec:U:1:SIID}

As another example we consider extending the Standard Model with both the complex singlet $S$ and the inert doublet $H_{2}$ \cite{Kadastik:2009dj,Kadastik:2009cu}. This is the minimal extension of the Standard Model that enables soft cubic couplings that can induce electroweak symmetry breaking \cite{Kadastik:2009ca,Kadastik:2009gx}, or semi-annihilation \cite{Belanger:2012vp}.

The necessary and sufficient vacuum stability conditions of the general $\Z_{2}$-symmetric model \cite{Kadastik:2009cu} can be derived via the Cottle-Habetler-Lemke theorem from the matrix of quartic couplings in the basis $(h_{1}^{2}, h_{2}^{2}, S_{H}^{2}, S_{A}^{2})$.%
\footnote{The conditions given in \cite{Kadastik:2009cu} are necessary conditions for positivity, not copositivity; in \cite{Kadastik:2011aa}  sufficient conditions were used.} However, to present a useful illustration of the copositivity criteria, we impose a global $U(1)$ symmetry on the potential.
With the given field content, the scalar potential invariant under the global $U(1)$ symmetry $H_{1} \to H_{1}$, $H_{2} \to e^{i \alpha} H_{2}$, $S \to e^{i \alpha} S$ transformations is
\begin{equation}
\begin{split}
  V &= \mu_{1}^{2} |H_{1}|^{2} +  \mu_{2}^{2} |H_{2}|^{2} + \mu_{S}^{2} |S|^{2} 
  + \frac{\mu_{SH}}{2} (\hc{H_{2}} H_{1} S + \hc{H_{1}} H_{2} \hc{S}) \\
  &+ \lambda_{S} |S|^{4} + \lambda_{1} |H_{1}|^{4} + \lambda_{2} |H_{2}|^{4} 
  + \lambda_{3} |H_{1}|^{2} |H_{2}|^{2} 
  + \lambda_{4} (H_{1}^{\dagger} H_{2}) (H_{2}^{\dagger} H_{1}) \\
  &+ \lambda_{S1} |S|^{2} |H_{1}|^{2} + \lambda_{S2} |S|^{2} |H_{2}|^{2}.
\end{split}
\label{eq:V:c}
\end{equation}
In the parameterisation \eqref{eq:V:2HDM:param}, the quartic part of the potential \eqref{eq:V:c} is
\begin{equation}
  V_{4} =
  \lambda_{S} s^{4} + \lambda_{1} h_{1}^{4}  + \lambda_{2} h_{2}^{4} 
  + \lambda_{3} h_{1}^{2} h_{2}^{2} + \lambda_{4} \rho^{2} h_{1}^{2} h_{2}^{2}
  + \lambda_{S1} h_{1}^{2} s^{2} + \lambda_{S2} h_{2}^{2} s^{2},
    \label{eq:V:Z5:1:and:2:SIID}
\end{equation}
where we have used $S = s e^{i \phi_{S}}$. The potential is minimised for $\rho = 0$ if $\lambda_{4} \geqslant 0$, and for $\rho = 1$ if $\lambda_{4} < 0$. As a shorthand for this, we can use the Heavyside theta function $\theta(-\lambda_{4})$. The matrix of couplings $\Lambda$ for the minimal potential in the $(h_{1}^{2}, h_{2}^{2}, s^{2})$ basis is given by
\begin{equation}
   2 \Lambda =
   \begin{pmatrix}
      2 \lambda_{1} & \lambda_{3} + \theta(-\lambda_{4}) \lambda_{4} & \lambda_{S1}
      \\
      \lambda_{3} + \theta(-\lambda_{4}) \lambda_{4} & 2 \lambda_{2} & \lambda_{S2}
      \\
      \lambda_{S1} & \lambda_{S2} & 2 \lambda_{S}
   \end{pmatrix}.
\end{equation}

Copositivity criteria \eqref{eq:A:3:copos:1:2:crit} and \eqref{eq:A:3:copos:3:crit} yield the necessary and sufficient vacuum stability conditions:
\begin{equation}
  \begin{split}
  \lambda_{1} &\geqslant 0, \quad \lambda_{2} \geqslant 0, \quad \lambda_{S} \geqslant 0, \\
  \bar{\lambda}_{12} &\equiv \lambda_{3} + \theta(-\lambda_{4}) \lambda_{4} 
  + 2 \sqrt{\lambda_{1} \lambda_{2}} \geqslant 0, \\
  \bar{\lambda}_{1S} &\equiv \lambda_{S1}  + 2 \sqrt{\lambda_{1} \lambda_{S}} \geqslant 0, \\
  \bar{\lambda}_{2S} &\equiv \lambda_{S2}  + 2 \sqrt{\lambda_{2} \lambda_{S}} \geqslant 0,
  \end{split}
  \label{eq::V:Z5:1:and:2:SIID:most}
\end{equation}
and
\begin{equation}
\begin{split}
  &\sqrt{\lambda_{1} \lambda_{2} \lambda_{S}} 
  + [ \lambda_{3} + \theta(-\lambda_{4}) \lambda_{4} ] \sqrt{\lambda_{S}} 
  + \lambda_{S1} \sqrt{\lambda_{2}} + \lambda_{S2} \sqrt{\lambda_{1}} 
  \\
  &+ \sqrt{ \bar{\lambda}_{12} \bar{\lambda}_{1S} \bar{\lambda}_{2S} }
  \geqslant 0.
  \end{split}
  \label{eq::V:Z5:1:and:2:SIID:last}
\end{equation}

Note that it is easy to take into account some terms that break the $U(1)$ symmetry. For example, we can add the $\lambda'_{S}$ term to the potential. As discussed in \secref{sec:SM:complex:singlet}, this corresponds to $\lambda_{S} \to \lambda_{S} - \abs{\lambda'_{S}}$ in \eqref{eq::V:Z5:1:and:2:SIID:most} and   \eqref{eq::V:Z5:1:and:2:SIID:last}. In the same way, we can add the $\lambda_5$ term, with $\lambda_{3} + \theta(-\lambda_{4}) \lambda_{4} \to \lambda_{3} + \theta(-\lambda_{4} + \abs{\lambda_{5}}) \, (\lambda_{4} - \abs{\lambda_{5}})$ in \eqref{eq::V:Z5:1:and:2:SIID:most} and   \eqref{eq::V:Z5:1:and:2:SIID:last}.

\section{Conclusions}
\label{sec:conclusions}

For a scalar potential whose $d = 4$ part is a biquadratic form of the fields, $\lambda_{ab} \varphi_{a}^{2} \varphi_{b}^{2}$, the vacuum stability conditions for the potential to be bounded below coincide with the criteria for the matrix of its quartic couplings $\lambda_{ab}$ to be copositive or positive on non-negative vectors. 
The notion of copositivity has found use in convex optimization, but has hitherto been overlooked in particle physics. Copositivity allows for a larger parameter space than positivity. For the latter, Sylvester's criterion gives a simple set of inequalities, but conditions for the former are more complicated. 

We reviewed basic properties of copositive matrices and explicit criteria for copositivity of $2 \times 2$ (these are already well known in particle physics literature, though not under that name) and of $3 \times 3$ matrices. The Cottle-Habetler-Lemke theorem or other algebraic criteria can be used for larger matrices. In code, the theorem is easier to use even for small matrices, as it is a polynomial criterion that does not involve calculation of square roots. For really large matrices (of about order $n > 20$), approximate algorithms have to be used. 

We illustrated this method by deriving the vacuum stability conditions for some dark matter models where the dark sector comprises either a complex singlet, an inert doublet, or both. In conclusion, copositivity allows to find analytic necessary and sufficient vacuum stability conditions for complicated models in an important special case.

\section*{Acknowledgements}

We thank Martti Raidal for comments and suggestions and Julia Polikarpus for consultation. This work was supported by the ESF grants 8090, 8943, MTT8, MTT60, MJD140 by the recurrent financing SF0690030s09 project and by the European Union through the European Regional Development Fund.

\bibliographystyle{JHEP}
\bibliography{copositivity}
\end{document}